\documentclass[pra,aps,twocolumn,showpacs,amsmath,amssymb]{revtex4}
\usepackage{graphicx}
\usepackage{dcolumn}
\usepackage{bm}
\usepackage{times}
\usepackage{epstopdf}
\usepackage{amsmath}
\usepackage{epsfig}
\newcommand{\ket}[1]{\left\vert#1\right\rangle}
\newcommand{\bra}[1]{\left\langle#1\right\vert}
\newcommand{\s}{\uparrow}
\newcommand{\g}{\downarrow}
\newcommand{\ug}{\!=\!}
\newcommand{\eq}{Eq.~}

\newcommand{\fig}{Fig.~}

\begin{document}
\author{F. Ciccarello\mbox{$^{1}$}}
\author{M. Paternostro\mbox{$^{2}$}}
\author{G. M. Palma\mbox{$^{3}$}}
\author{M. Zarcone\mbox{$^{1}$}}
\affiliation{\mbox{$^{1}$}CNISM and Dipartimento di Fisica e
Tecnologie Relative, Universit\`{a} degli Studi di Palermo, Viale
delle Scienze, Edificio 18, I-90128 Palermo, Italy \\
\mbox{$^{2}$} School of Mathematics and Physics, Queen's
University, Belfast BT7 1NN, United Kingdom\\
\mbox{$^{3}$} NEST-INFM (CNR) , and Dipartimento di Scienze
Fisiche ed Astronomiche, Universit\`{a}
degli Studi di Palermo, Via Archirafi 36, I-90123 Palermo, Italy}

\begin{abstract}

We present a protocol that sets maximum stationary entanglement between remote spins through scattering of mobile mediators without initialization, post-selection or feedback of the mediators' state. No time-resolved tuning is needed and, counterintuitively, the protocol generates two-qubit singlet states even when classical mediators are used. The mechanism responsible for such effect is resilient against non-optimal coupling strengths and dephasing affecting the spins.  The scheme uses itinerant particles and scattering centres and can be implemented in various settings. When quantum dots and photons are used a striking result is found: injection of classical mediators, rather than quantum ones, improves the scheme efficiency.

 \end{abstract}


\title{Reducing quantum control for spin-spin entanglement distribution}
\maketitle

Enforcing a state in a quantum system, a task usually requiring quantum control~\cite{Varie0,Varie0bis}, is key to the grounding of quantum technology~\cite{NC}. Typically, state initialisation, interaction-tuning, postselection and feedback are needed in order to achieve a given state~\cite{proposals,proposals1,Varie0tris,Varie1,Varie2,Varie3,Varie4,feed1,feed2,postsel,postsel2}, especially when a system involves remote parties requiring interaction-bridging mediators~\cite{distant0,distant1,distant2,distant3,distant4,distant5}. One would expect that quantum coherence in the state of such mediators is needed to make two remote particles interact. Here, we discuss a protocol that sets maximum stationary entanglement between remote spins without initialization of mediators' state, post-selection or feedback. No time-resolved tuning is needed and, counterintuitively, the protocol generates two-qubit singlet states even when classical mediators are used. The mechanism behind this process is stable against non-optimal coupling strengths and robust against dephasing affecting the spins. Our proposal uses flying particles (conduction electrons or photons) and two scattering centres (substitutional impurities in nanowires or quantum dots in waveguides).

\begin{figure}[t]
\psfig{figure=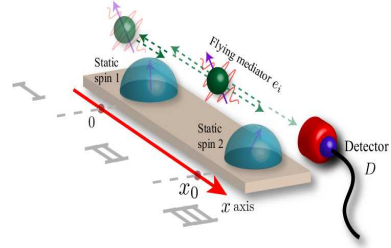,width=5.cm,height=3.4cm}
\caption{General set-up for the proposed scheme. Two static spins, encoded in the bi-dimensional space $\{\ket{\uparrow,\downarrow}\}$, separated by a distance $x_0$, interact with a stream of flying particles $e_n$'s, each being a spin-$1/2$ prepared in a {\it classical} statistical mixture $(1/2)(\ket{\uparrow}\bra{\uparrow}+\ket{\downarrow}\bra{\downarrow})$. While in general each $e_n$, after multiple scattering between 1 and 2, can be reflected back, we collect at $D$ only the particles that successfully trespass the interaction region of $0\le{x}\le{x}_0$.}
\label{fig1}
\end{figure}

In order to best present these ideas, we use a setup-independent language that lets us stress the flexibility of our mechanism. We consider two static spin-$1/2$ particles, labelled $1$ and $2$, separated by a distance $x_0$ in a one-dimensional structure, as in Fig.~\ref{fig1}. 
A low-density stream of mobile spin-$1/2$ particles, $e_n$'s, propagates along $x$: each of them 
undergoes scattering by particles $1$ and $2$ whenever at 
their respective sites. In our notation, each (static or mobile) spin is encoded in the bi-dimensional space $\{\ket{\uparrow,\downarrow}\}$. The interaction Hamiltonian ruling the bilocal coupling between 
each mobile spin $e_n$ and the static ones is
\begin{equation} 
\label{H}
\hat{H}_n={\hat{p}_n^{2}}/({2m})\!+\! J \,
\hat{\mbox{\boldmath$\sigma$}}_{n}\cdot
[\hat{\mathbf{S}}_1\,\delta(x)+
\hat{\mathbf{S}}_2\,\delta(x-x_0)],
\end{equation}
where $\hat{p}_n$ ($m$) is the momentum operator (mass) of $e_n$, $J$ 
is the interaction strength and $\hat{\mbox{\boldmath$\sigma$}}_n$,
$\hat{\mathbf{S}}_1$ and $\hat{\mathbf{S}}_2$ are the spin 
operators of $e_n$, $1$ and $2$, respectively (throughout this paper we adopt units such that $\hbar\ug1$). Our first task is to  demonstrate that, under proper geometric conditions on this general setup and by simply requiring the conservation of the number of mobile particles crossing the scattering region, maximum stationary entanglement \cite{NC} between $1$ and $2$ can be set, regardless of the state the $e_n$'s are prepared in. 
We start examining the general symmetries enjoyed by Hamiltonian (\ref{H}). These arise from the commutation rules  $[\hat{H}_n,\hat{\mathbf{S}}^2_n]=[\hat{H}_n,\hat{S}_{z,n}]=0$, where $\hat{\mathbf{S}}_n\ug \hat{\mbox{\boldmath$\sigma$}}_n+\hat{\mathbf{S}}_{12}$ is the total spin operator of the $e_n\!-\!1\!-\!2$ system and $\hat{\mathbf{S}}_{12}=\hat{\mathbf{S}}_1+\hat{\mathbf{S}}_2$.
For each left-incoming $e_n$, the overall spin space is eight-dimensional. Moreover, the only free-energy term in~\eq(\ref{H}) is the kinetic one so that, for an $e_n$ injected with wavevector  $k_n$, the system's energy is $E_{k_n}\ug k^2_n/(2m)$ with eight associate (degenerate) stationary states, each corresponding to a different spin state. Knowledge of their form allows us to determine the evolution of the system's initial spin-state upon each scattering event. It is straightforward to prove (cfr. Appendix A) that if $k_n x_0\ug q \pi$ (with $q$ a positive integer),  a situation which we call ``resonance condition" (RC), dynamics takes place as if particles $1$ and $2$ occupy the same site. We have the formula $\delta_{RC}(x)\ug\delta_{RC}(x-x_0)$ (the subscripts remind us of the RC condition). This effect has a clear interpretation related to the phase-factors $e^{\pm i k_n x_0}$ acquired by $e_n$'s wavefunction upon multiple reflections at the sites of particles $1$ and $2$ (cfr.~\fig 1). Under RC $e^{\pm i k_n x_0}\ug1$ and no relative phase-shift occurs between reflected and transmitted components of the wavefunction, as if the scattering centers were at the same site. This introduces additional symmetry to the effective interaction between 1 and 2 mediated by $e_n$, as witnessed by the new conservation law
$[\hat{H}_n,\hat{\mathbf{S}}_{12}^2]=0$. This arises simply by noticing that under RC 
$k_n\ug k_{RC}\ug q \pi /x_0$ and 
\eq (\ref{H}) takes the effective form
\begin{equation} \label{HRC}
\hat{H}_{RC,n}=q^2\pi^2
/({2mx_0^2})\!+\! J/2 \,
(\hat{\mathbf{S}}_n^2-\hat{\mbox{\boldmath$\sigma$}}^2_n-
\hat{\mathbf{S}}_{12}^2)\,{\delta}_{RC}(x),
\end{equation}
while $\hat{\mathbf{S}}_{12}^2$ commutes with each term in \eq (\ref{HRC}).
\begin{figure}[t]
\psfig{figure=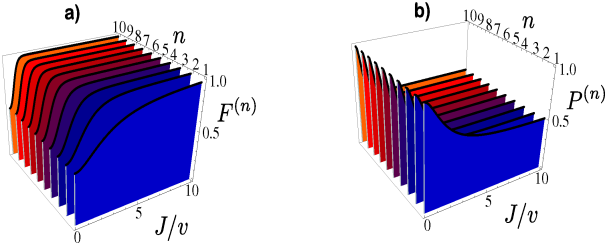,width=8.0cm,height=3.2cm}
\caption{Performance of the scheme.  (a) State fidelity $F^{(n)}\ug\!_{12}\!\langle\Psi^{-}|\rho_{12}^{(n)}|\Psi^{-}\rangle_{12}$ between the target state $|{\Psi^-}\rangle_{12}={2}^{-1/2}(\ket{\uparrow\downarrow}-\ket{\downarrow\uparrow})_{12}$ and the remote-spins' state $\rho_{12}^{(n)}$ obtained after $n$ mediators have been injected and all counted at $D$. (b) Associated probability of success $P^{(n)}$. We use the Hamiltonian model in Eq.~(\ref{HRC}) and call $v\ug k/m$ the mediator velocity (assuming all the mediators to have the same wavevector $k$). The curves are insensitive to the spin state of each $e_n$, which can therefore be even unknown.} 
\label{fig0NP}
\end{figure}
We use the state fidelity $F^{(n)}\ug\!_{12}\!\langle\Psi^{-}|\rho_{12}^{(n)}|\Psi^{-}\rangle_{12}$.  In Fig.~\ref{fig0NP}(a), we consider the initial product state $\rho_{12}\ug\ket{\s,\g}_{12}\!\langle\s,\g\!|$ and plot $F^{(n)}$ against $n$ and the ratio $J/v$ ($v\ug k/m$, assuming all the mediators to have the same wavevector $k$).

The above features are enough to explain the claimed insensitivity of entanglement generation to the mediators' internal state. Let us consider an initial spin state of the form $\ket{\chi}_{e_n}\! \ket{\Psi^-}_{12}$, where $\ket{\chi}_{e_n}$ is an \textit{arbitrary} spin state of $e_n$ and
 $\ket{\Psi^-}_{12}\ug 2^{-1/2}(\ket{\s\g}_{12}-\ket{\g\s}_{12})$ is the maximally entangled singlet state of 1 and 2.  $\ket{\chi}_{e_n} \ket{\Psi^-}_{12}$ is an eigenstate of $\hat{\mathbf{S}}_{12}^2 $ with zero eigenvalue.
On the other hand, $\ket{\chi}_{e_n} \ket{\Psi^-}_{12}$ is also an eigenstate of $\hat{\mathbf{S}}_n^2$  with eigenvalue $3/4$, which brings us directly to the very special effective Hamiltonian $\hat{H}_{RC,n}\ug q^2\pi^2/({2mx_0^2})$, where the spin degrees of freedom of particles $e_n$, $1$ and $2$ are absent. Despite its simplicity, this result brings about two crucial consequences. First,  each $e_n$ is transmitted through the interaction region with $100\%$ probability whenever 1 and 2 are in the singlet state. That is, by placing a Geiger-like particle counter $D$ at the right-hand side of the setup (as in Fig.~\ref{fig1}), the number of flying mediators trespassing 1 and 2 is conserved. Second, each initial spin state  $\ket{\chi}_{e_n}\!\ket{\Psi^-}_{12}$ is left unchanged by the scattering dynamics (remind that $\ket{\chi}_{e_n}$ is arbitrary). This property is easily extended to arbitrary mixed spin states of each $e_n$. The singlet state of spins $1$ and $2$ is the {\it only} initial state to enjoy such features, as can be easily proved by considering simple properties of addition of angular momenta. Also, notice that our arguments are valid for any coupling strength $J$.

In retrospect, if in an experiment the number of particles passing through the interaction region is conserved (that is, $D$ counts as many clicks as the number of particles that have been injected) in virtue of the previous analysis we conclude that, after a sufficiently large number of mediators' injections particles $1$ and $2$ necessarily are in a singlet state. Our proposal de facto embodies a Bell-like projective measurement~\cite{NC} performed over the remote spins. This occurs regardless of the spin-state of each $e_n$, which can be prepared even in the classical statistical mixture $\rho_{e_n}\ug1/2( \ket{\g}_{e_n}\! \bra{\g}\!+\!\ket{\s}_{e_n}\! \bra{\s})\ug{I}_{e_n}/2$, therefore demonstrating that maximum entanglement can be distributed between two remote spins with virtually no quantum control needed on the state of each mediator. We now show that the scheme's efficiency depends on parameters that can be engineered off-line.

 We start by clarifying how many mediators one needs  to inject and count at $D$ before projection onto $\ket{\Psi^-}_{12}$ is achieved with significant fidelity. This depends on the coupling strength $J$. 
 In fact, for a given state 
 $\rho_{12}\!\neq\! \ket{\Psi^-}_{12}\! \bra{\Psi^-}$, each $e_n$ has a non-null probability to be reflected back without reaching $D$. Such a reflection probability  grows with $J$, while we already know it is zero if $\rho_{12}\!\ug\! \ket{\Psi^-}_{12}\! \bra{\Psi^-}$, {regardless} of $J$. A large value of $J$ thus makes us confident that only a few transmitted mediators need to be counted at $D$ before the effective Bell-projection is achieved. Indeed, in this case, we will be confident that the transmission of all $e_n$'s is associated with particles $1$ and $2$ being in a state very close to $\ket{\Psi^-}_{12}$. We remark that, although the number of required mediators depends on the coupling strength, the convergence to the singlet is asymptotically achieved for \textit{any} value of this parameter. No fine setting of $J$ is required by our scheme.
 Furthermore, in line with the features of scattering-based protocols~\cite{distant3}, only a very weak requirement on the control of interaction times is in order:  the time elapsed between the injection of two successive mediators should exceed the characteristic time $T_s$ taken by each scattering process. Under easily-matched conditions and for quasi-monochromatic Gaussian wavepackets of the mediators \cite{rising-time}  we have $T_s\!\sim\!1/(v_{k_0}\Delta k)$ with $v_{k_0}$ the velocity associated with the carrier wavevector $k_0$ and $\Delta k$ each wavepacket-width in $k$-space. This tells us that the characteristic time taken by each scattering event depends \textit{only} on kinetic parameters of the mediator wavepacket and not on the spin-spin coupling mechanism. The mediators' wavepackets can be taken so as to make $T_s\ll{T}_{d}$, the latter being the characteristic time of coherent dynamics of a given setup (i.e. the time-scale after which decoherence affecting spins 1 and 2 yields significant effects). 
 
 While the robustness of the protocol presented here to dephasing affecting the static spins is discussed later on, in Fig.~\ref{fig0NP}  we study the efficiency of the scheme. We consider the initial product state $\rho_{12}\ug\ket{\s,\g}_{12}\!\langle\s,\g\!|$ and use the fidelity $F^{(n)}$ to measure how close to $|\Psi^{-}\rangle_{12}$ is the 1-2 state after that $n$ successive mediators have been injected and have all been detected at $D$. We call $P^{(n)}$ the corresponding success probability.
 In agreement with our explanations, $F^{(n)}$ grows with $n$ and progressively approaches $F^{(n)}\ug1$, which marks the generation of a singlet state. On the other hand, Fig.~\ref{fig0NP}(b) shows how the probability $P^{(n)}$ that all the mediators are collected at $D$  converges to $1/2$. The curves in these plots are insensitive to the degree of purity of the state of each $e_n$ (cfr. Appendix A). In particular, as we stated above, these same features hold for an injected stream of particles each prepared in the unpolarized state $\rho_{e_n}=I_{e_n}/2$. This classical statistical mixture gives us the least possible information on each mediator's spin state. Notice that the rate of convergence of both $F^{(n)}$ and $P^{(n)}$ increases with $J/v$ ($v$ is the mediator velocity), in agreement with our predictions: the stronger the spin-spin coupling, the smaller the required $n$.
For instance, at $J/v\simeq 1.6$,  $F^{(5)}\!>\!95\%$ with $P^{(5)}\!>\!50\%$, while at $J/v\!\gtrsim \!7.5$ even a \textit{single} mediator being sent and collected is enough to achieve these values.

An interesting arena for a solid-state implementation of the protocol described so far is served by a one-dimensional (1D) CdTe nanowire. A stream of conduction electrons would embody particles $e_n$'s and substitutional Mn atoms could act as scattering centers 1 and 2. In the CdTe host, Mn atoms are not ionized due to valence-number matching, which suppresses any mediator-static spin electrostatic interaction and leaves only an exchange-type spin-spin coupling. The conditions of  extremely low control required by our scheme fit very well with the experimental capabilities in current spintronics settings such as the one sketched here. As no preparation or post-selection of the mediator's spin-state is required, there is no need for experimentally challenging spin-filtering operations performed over mobile electrons (the efficiency of currently available spin filters is to the best of our knowledge still quite low~\cite{spin-filter}). On the other hand, the fabrication of Mn-doped nanostructures is currently experiencing some impressive improvements, including the implantation of a {single} Mn atom into a quantum dot~\cite{grenoble}. Although a Mn atom in a CdTe compound has quantum spin number $s=5/2$, while the description of our proposal involved spin-$1/2$ particles, the working principle of our scheme remains valid, with due adjustments, in the case of spin-$s$ scattering particles (cfr. Appendix A). 

\begin{figure}[t]
\psfig{figure=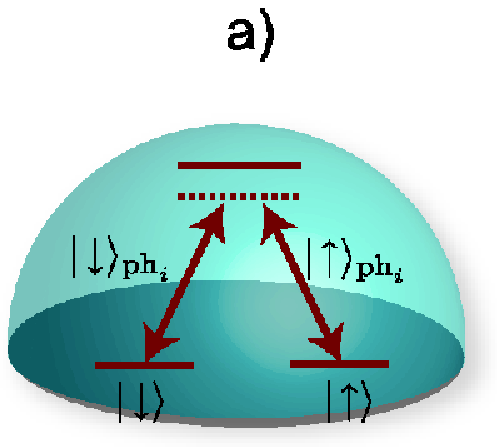,width=2.2cm,height=1.8cm}\psfig{figure=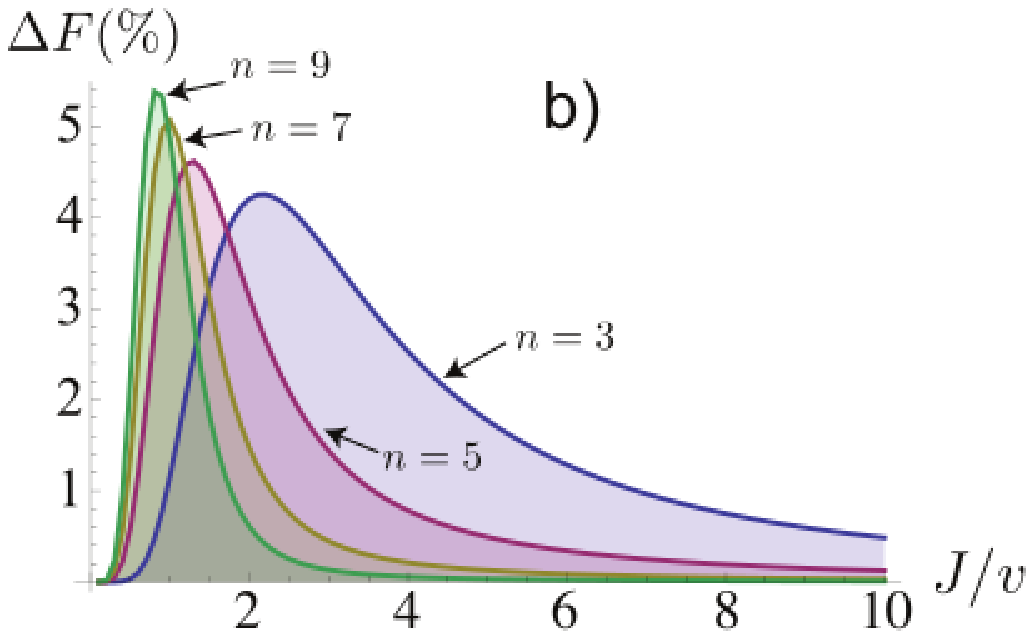,width=6.0cm,height=3.6cm}
\caption{Flexibility of the scheme to alternative coupling models. (a) Level-configuration of the effective static pseudospin-$1/2$ used in a cavity-quantum electrodynamics implementation of our scheme. 
(b) Percentage difference between the state fidelity $F^{(n)}$ achieved upon use of models $\hat{H}_{RC,n}\!=\!q^2\pi^2/({2mx_0^2})\!+\! J/2 \,(\hat{\mathbf{S}}_n^2-\hat{\mbox{\boldmath$\sigma$}}^2_n-\hat{\mathbf{S}}_{12}^2)\,{\delta}_{RC}(x)$ and $\hat{\mathcal{H}}_{RC,n}\!=\!q\pi{v}_{\text{ph}}/x_0\!+\!J\,[\hat{\sigma}_{-,n}(0)\hat{S}_{+,12}+{\rm h.c.}],$ against  the rescaled spin-spin interaction strength and for a few significant values of $n$. Results quantitatively very close to those shown here are found for the percentage differences in success probability $P^{(n)}$.} 
\label{fig1NP}
\end{figure}

\begin{figure}[t]
\psfig{figure=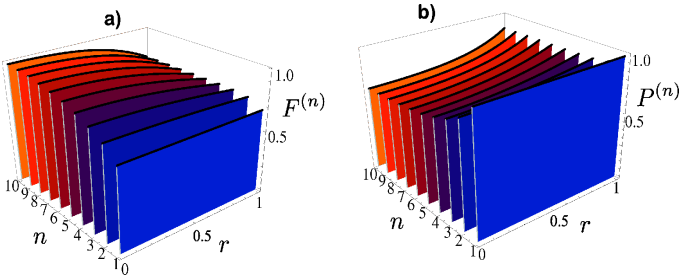,width=9.0cm,height=3.6cm}
\caption{Effects of knowledge of mediators' state. (a) Fidelity and (b) success probability of the proposed scheme, when the model valid in the cavity-quantum electrodynamics implementation is used, against $n$ and the parameter $r$ entering the state of the injected mediators. For  $r\ug0$ we prepare initial statistical mixtures with no quantum coherence, which nevertheless correspond to the optimized effective Bell-projection described in the text. } 
\label{fig2NP}
\end{figure}

A second interesting scenario for testing our predictions is provided by a cavity-quantum electrodynamics (QED) setup. Our proposal 
considers  a 1D semiconductor photonic waveguide offering the advantages of reliable photonic transport. Two embedded multi-level quantum dots (QDs) embody two effective pseudospin-$1/2$ static particles. The spectrum of each dot consists of a ground doublet and one excited state. The waveguide accommodates two frequency-degenerate, orthogonally-polarised modes of radiation. Each mediator $e_n$ and its spin states 
are respectively embodied by a photon $\text{ph}_n$ and its (orthogonal) polarization states, which we abstractly indicated as $\ket{\g,\s}_{\text{ph}_n}$ [cfr. Fig.~\ref{fig1NP}(a)]. 
For a large enough detuning between each pseudospin transition-frequency and the photonic mediators, the excited state is only virtually populated and 
the transition between the pseudospin states is achieved via two-photon Raman
processes with associated coherent scattering of a photon between
states $\ket{\g}_{\text{ph}_n}$ and $\ket{\s}_{\text{ph}_n}$. We consider a linear photonic dispersion relation $E_{k_n}\ug v_{\text{ph}} k_n$ for each propagating photon so that the Hamiltonian describing the dynamics of QD's and each photonic mediator under RC takes the effective form~\cite{distant0}
\begin{equation}
\label{HRC-ph}
\hat{\mathcal{ H}}_{RC,n}\ug q\pi{v}/x_0\!+\!J\,\left[\hat{\sigma}_{-,n}(0)\hat{S}_{+,12}+{\rm h.c.}\right],
\end{equation}
where $\hat{\sigma}_{+,n}(x)\ug\hat{\sigma}^\dag_{-,n}(x)$ are effective pseudospin density photon operators at position $x$ (cfr. Appendix A) whereas $\hat{S}_{\pm,12}\ug\hat{S}_{\pm,1}\!+\!\hat{S}_{\pm,2}$  and $S_{+,i}\ug S^\dag_{-,i}$ is the ladder operator of QD $i\!=\!1,2$. With due changes, the approach followed in order to demonstrate our scheme holds under the model in Eq.~(\ref{HRC-ph}). 
The results corresponding to this case are indeed very similar to those reported in Fig.~{\ref{fig0NP}}. In fact, both $F^{(n)}$ and $P^{(n)}$
corresponding to model (\ref{HRC-ph}) deviate by less than $6\%$, at small values of $J$, from the analogous quantities in the case of Eq.~(\ref{HRC}). The difference drops to zero in the range of parameters that guarantee high efficiency of the effective Bell-projection. Fig.~\ref{fig1NP}(b) shows the percentage difference between such quantities, which explicitly shows the closeness of the two models. The results in Fig.~\ref{fig1NP} are obtained for a stream of unpolarized photonic mediators, each in $\rho_{\text{ph}_n}\ug I_{\text{ph}_n}/2$. This brings us to the next question: how does the state of each mediator affect the performance of the scheme for the model in \eq (\ref{HRC-ph}) [we already know that the efficiency obtained through \eq(\ref{HRC}) is insensitive to the mediators' spin state]?
To assess this, we consider the case that each mediating photon is injected in the state $\rho_{\text{ph}_n}\ug [(1-r) \ket{\g}_{\text{ph}_n}\! \bra{\g}\!+\!(1+r) \ket{\s}_{\text{ph}_n}\! \bra{\s}]/2$, where $r\in[0,1]$  determines the purity of the polarization state. 
In~Fig.~\ref{fig2NP} we set $J/v\ug 1.5$ and plot $F^{(n)}$ and $P^{(n)}$ against $n$ and $r$. Surprisingly, the efficiency of the scheme \textit{decreases with} $r$: the distribution of maximum entanglement is \textit{optimized} by sending fully unpolarized photons. Despite its counter-intuitive nature, this result is easily understood by noticing that, in full analogy with the model in Eq.~(\ref{HRC}), if the pseudospins are in a singlet state, the spin-spin term in~\eq (\ref{HRC-ph}) vanishes \textit{regardless} of the mediators' state, implying that the injected photons will all be collected at $D$. However, unlike the case of an exchange interaction, as in Eq.~(\ref{H}), this is not the only case where such behaviour occurs. Indeed, for an incoming photon in  $\rho_{\text{ph}_n}\ug \ket{\s}_{\text{ph}_n}\! \bra{\s}$ ($r\!=\!1$) with the pseudospins in $\rho_{12}\ug|\!\s,\!\s \rangle_{12}\langle \s,\!\s\!|$, it is $\hat{\mathcal{ H}}_{RC,n}=v_{\text{ph}}k_n$ (for any $k_n$). However, when $r\!\neq\!1$ this is no longer true since the photon has now some probability to be sent in $|\!\g \rangle_{\text{ph}_n}\!\langle\g\!|$. In this case, the previous argument does not hold and some ``impedance" to the mediator's transit arises even when $\rho_{12}\ug|\!\s,\!\s \rangle_{12}\!\langle\!\s,\!\s\!|$:  the higher $r$, the stronger the impediment. As the latter is the basic resource harnessed by our scheme in order to discriminate the singlet state, our analysis explains and clarifies why the protocol is optimized when $r\!\ug\!0$.

We have shown the possibility of guiding the state of a distributed system of remote spins via a control-relaxed scattering. Beside proper geometrical arrangement of the setup, implying a properly set inter-spin distance, the mere conservation of the number of scattered mediators is required for the success of the protocol. Preparation and post-selection of the mediators' internal state are not demanded by our scheme, which sets steady-state entanglement insensitive, by definition, to timing imperfection. Against any expectations, even \textit{classical} mediators can be used to establish maximum entanglement. We have demonstrated that the scheme is stable against non-optimal coupling strengths while it exhibits an intrinsic robustness against collective dephasing-like noise. In fact, our effective Bell projection progressively extracts the singlet-state component from the initial state of the remote spins, and it is well-known that collective scrambling mechanisms affecting the phase-relation within the state of the distributed spins is quenched by the symmetries of the singlet state \cite{massimoProcRoy}. The quantitative aspects of this study, together with those relative to the case of individual noise-mechanisms affecting the spins, are presented in Appendix B. Within the context of quantum-state guidance, our proposal stems as a milestone demonstrating that full advantage can be taken from suitable symmetries in order to significantly reduce quantum control. 

\acknowledgments

We thank A. Acin, G. Fishman and J. M. Lourtioz for discussions. We acknowledge support from PRIN 2006 ``Quantum noise in mesoscopic systems'', EUROTECH and the British Council/MIUR British-Italian Partnership Programme 2007-2008. MP is supported by EPSRC (EP/G004579/1).\\

{\it Note added.} After completion of the manuscript we became aware of a related proposal put forward by K. Yuasa in arXiv:0908.4377v1 [quant-ph].

\renewcommand{\theequation}{A-\arabic{equation}}
\setcounter{equation}{0}  
\section*{APPENDIX A: Technical tools}  
\label{AppTec}

Throughout these Appendices we omit the subscript $n$ in the scattering mediator since the features discussed here do not depend on this label.
 
 \subsection*{1. Stationary states} 
Here, we give a brief account of the method followed in order to determine the stationary states corresponding to the generalization of the interaction models in the body of the paper  to remote static particles having spin quantum number $s\ge{1/2}$. We call $m_{e}\!=\g,\s$ and $m_{i}\!\ug -s,..s$ the quantum numbers of $\sigma_{z}$ and $S_{iz}$ respectively ($i\ug1,2$). 
A spin state $|m_{e},m_1,m_2\rangle$ is labeled by $\mu\ug\{\!m_{e}\!,m_1\!,m_2\!\}$. We consider the three sections into which the $x$ axis in Fig.~1 has been divided. The system's stationary states can be written as $|\!\Psi^{\mu'}\!\rangle\ug\sum_{\mu}\int dx\,\Psi^{\mu'}_\mu(x)|\mu\rangle|x\rangle$ with the wavefunctions $\Psi_{\mu,\!I\!}^{\mu'}(x)\ug \delta_{\mu,\mu'}e^{ikx}\!+\!r_{\mu}^{\mu'}e^{-ikx}$, $\Psi_{\mu,\!I\!I\!}^{\mu'}(x)\ug A_{\mu}^{\mu'}e^{ikx}\!+\!B_{\mu}^{\mu'}e^{-ikx}$ and $\Psi_{\mu,\!I\!I\!I}^{\mu'}(x)\ug t_{\mu}^{\mu'}e^{ikx}$, each valid in section I, II and III respectively. Here, coefficients $A$'s, $B$'s, $r$'s and $t$'s are computed by imposing proper boundary conditions \cite{ciccarello2} on $\Psi_{\mu}^{\mu'}(x)$ and, depending on the interaction model, its derivative at  $x\ug0$ and $x\ug x_0$. It should be remarked that $\Psi_{\mu}^{\mu'}(x)$ as well as $A$, $B$, $r$ and $t$ implicitly depend on $k x_0$ and $J/v$.

 \subsection*{2. On the consequences of resonance conditions}
 
The previous section allows us to prove that under resonance condition (RC)  ${\delta}_{RC}(x)\ug{\delta}_{RC}(x\!-\!x_0)$. In the basis $\{|\!\Psi^{\mu'}\rangle\}$ the matrix representation of the operator ${\delta}(x\!-\!x')$ $\mathbf{ M}({x'}\!)$ is easily found to be $M\!(x'\!)_{\!\mu'',\mu'\!}\ug \langle \Psi^{\mu''}\!|{\delta}(x\!-\!x')|\Psi^{\mu'}\rangle\ug \sum_{\mu}\bar{\Psi}_{\mu}^{\mu''}\!(x')\Psi_{\mu}^{\mu'}\!(x')$ ($\bar{z}$ is the complex conjugate of $z$).  Matching of $\Psi_{\mu}^{\mu'}(x)$ at $x\ug x_0$ implies that $A_{\mu}^{\mu'}\!+\!B_{\mu}^{\mu'}e^{-2ikx_0}\ug t_{\mu}^{\mu'}$, which for $k x_0\ug q \pi$ becomes $\Psi_{\mu}^{\mu'}(0)\ug e^{-iq\pi}\Psi_{\mu}^{\mu'}(x_0\!)$. This gives $\mathbf{ M}({0})\ug \mathbf{ M}({x_0}\!)$ and thus ${\delta}_{RC}(x)\ug {\delta}_{RC}(x\!-\!x_0)$.

 \subsection*{3. Description of the spin state after a scattering event} 
 Here we assess the form taken by the spin state of the remote spins after scattering of a mediator takes place. Let $\rho\!=\!\rho_{e}\rho_{12}$ be the initial spin state. Its decomposition in the spin basis $\{\ket{\mu}\}$ introduced above is $\rho\ug\sum_{\mu,\mu'}c_{\mu,\mu'}\!|\mu\rangle\langle \mu'\!|$. The corresponding scattering process is fully described by the state $\rho_s\ug\sum_{\mu,\mu'}c_{\mu,\mu'}\!|\Psi^\mu\rangle\langle \Psi^{\mu'}\!|$ (notice that, unlike $\rho$, $\rho_s$ refers to both the spatial and spin degrees of freedom). The transmitted part of $\rho_s$ is $\rho_t(x,x')\ug\sum_{\mu,\mu'}c_{\mu,\mu'}\sum_{\nu}t_{\nu}^{\mu}e^{ikx}|\mu\rangle\,\sum_{\nu'}\!\bar{t}_{\nu'}^{\mu'}e^{-ikx'}\langle \nu'|$, where $t^{\mu}_{\nu}$'s are transmission amplitudes associated with scattering between spin states $\ket{\mu}$ and $\ket{\nu}$. The reflected part can be determined, likewise, in terms of analogous reflection amplitudes $r^{\mu}_{\nu}$'s. Once trace of $\rho_s$ over $e$ is performed, the final state of the remote spins depends on operators $\hat{T}_{m_e}^{m'_e}$'s and $\hat{R}_{m_e}^{m'_e}$'s, which are functions of the transmission and reflection amplitudes $t$'s and $r$'s, respectively (cfr. following Section). Their analytical expressions are too cumbersome to be presented here.

\subsection*{4. Effect of mediator counts} 
In order to give a quantitative account of how recording clicks at the Geiger-like detector $D$ (shown in Fig.~1) affects the remote-spin state we adopt the language of quantum maps~\cite{NC}. Assume that  a mediating particle $e$ has been prepared in the state $\ket{m'_e\!=\g,\s}$ and sent to the interaction region, while spins 1 and 2 have been prepared in $\rho_{12}$. Without the Geiger-like detector, the state of 1 and 2 after scattering is
\begin{equation}
\label{map_gen}
\rho'_{12}=\sum_{m_e\ug\s,\g}(\hat{R}^{m'_e}_{m_e}\rho_{12}\hat{R}_{m_e}^{m'_e\,\dag}+\hat{T}^{m'_e}_{m_e}\rho_{12}\hat{T}_{m_e}^{m'_e\,\dag}),
\end{equation}
where $\hat{R}^{m'_e}_{m_e}$'s ($\hat{T}^{m'_e}_{m_e}$'s) are the operators (introduced in the previous Section) describing how a reflected (transmitted) mediator in $\ket{m_e\ug\g,\s}$ affects the static spins' state. 
In the case that $e$ is injected in a state of the form $\rho_e\ug \ug r \ket{\g}_e\! \bra{\g}\!+\!(1\!-\!r) \ket{\s}_e\! \bra{\s}$, recording a click at $D$ changes the initial state $\rho_{12}$ into
\begin{equation}\label{map_up}
\mathcal{\tilde{E}}^{(1)}(\rho_{12})\ug r \!\sum_{m_e\ug\s,\g}\hat{T}^{\g}_{m_e}\rho_{12}\hat{T}_{m_e}^{\g\,\dag}\!+(1\!-\!r)\!\sum_{m_e\ug\s,\g}\hat{T}^{\s}_{m_e}\rho_{12}\hat{T}_{m_e}^{\s\,\dag}.
\end{equation}
As, in general, $e$ may also be reflected back, state (\ref{map_up}) is not normalized. Thus, the right-hand side  has to be divided by the probability $P^{(1)}(\rho_{12})={\text{Tr}}[\mathcal{\tilde{E}}^{(1)}(\rho_{12})]$ that $e$ is transmitted. Therefore, the complete map transforming $\rho_{12}$ into the static spins' state $\rho^{(1)}_{12}$ after the injection and collection at $D$ of a single mediator reads $\rho^{(1)}_{12}\!=\!\mathcal{E}^{(1)}(\rho_{12})\ug\mathcal{\tilde{E}}^{(1)}(\rho_{12})/P^{(1)}(\rho_{12})$. The state $\rho^{(n)}_{12}$ after $n$ of such injection-collection steps is given by the $n$-time application of the map as $\rho^{(n)}_{12}\ug \mathcal{E}^{(n)}(\rho_{12})\ug\mathcal{E}[\mathcal{E}[\cdot\cdot[\mathcal{E}(\rho_{12})]]]$ with associated probability $P^{(n)}\!=\!{\text{Tr}}[\mathcal{\tilde{E}}^{(n)}(\rho_{12})]$, where $\mathcal{\tilde{E}}^{(n)}(\rho_{12})]$ is obviously defined as the $n$-time application of the map in Eq.~(\ref{map_up}) above.
As discussed in the body of the paper, when $\rho_{12}\ug|\Psi^-_{12}\rangle \bra{\Psi^{-}_{12}}$ $e$ is always collected at $D$ and, in addition, the remote spins' state is not affected. Thus, under RC $\mathcal{E}^{(n)} (|\Psi^-_{12}\rangle \bra{\Psi^{-}_{12}})\ug|\Psi^-_{12}\rangle \bra{\Psi^{-}_{12}}$ and $P^{(n)} (|\Psi^-_{12}\rangle \bra{\Psi^{-}_{12}})\ug1$ for any $n$, $r$ and $J$. In the case of the Hamiltonian given in Eq.~(2), this is 
the only spin state enjoying such property, which characterizes the singlet as the \textit{only fixed point} of map $\mathcal{E}$. For Eq.~(3), on the other hand, it is the only fixed point provided that $r\neq{0}$.

\subsection*{5.  Cavity-quantum-electrodynamics model}  

In this Section we provide a few key details on the derivation of the effective Hamiltonian Eq.~(3).  For a generic photon wavevector, we have $\hat{\mathcal{H}}\ug \hat{\mathcal{H}}_0+\hat{V}$, where $ \hat{\mathcal{H}}_0$ is the free-Hamiltonian of the waveguide and $\hat{V}$ is the atom-photon interaction. $ \hat{\mathcal{H}}_0$  reads
\begin{equation}
\label{H0-ph}
\hat{\mathcal{H}}_{0}=-i\sum_{\beta=R,L}\sum_{\gamma=\s,\g}\int dx\,
v_{\beta}\,\hat{c}_{\beta,\gamma}^\dag(x){\partial}_x\hat{c}_{\beta,
\gamma}(x),
\end{equation}
where $v_{R}\ug-v_L\ug v$ and $\hat{c}_{R,\gamma}^\dag(x)$ [$\hat{c}_{L,\gamma}^\dag(x)$] is the bosonic operator creating a right (left) propagating photon of
polarization $\gamma\ug\s,\g$ at position $x$. $\hat{V}$ has the form
\begin{equation}
\label{api }
\hat{V}\ug\int{d}x\,\sigma_{+}(x)[\hat{S}_{1-}\,\delta(x)\!+\hat{S}_{2-}\delta(x-x_0)]\,+\mathrm{h.c.}\,\,,
\end{equation}
where $\hat{\sigma}_{+}(x)\ug \hat{\sigma}_{-}^{\dag}(x)\ug \hat{c}_{\s}^\dag(x)\hat{c}_{\g}(x)$, along with $\hat{\sigma}_z(x)=[\hat{c}_{\s}^\dag(x)\hat{c}_{\s}(x)-\hat{c}_{\g}^\dag(x)\hat{c}_{\g}(x)]/2$, are pseudo-spin density operators such that $\hat{\mbox{\boldmath$\sigma$}}=\int
dx\,\hat{\mbox{\boldmath$\sigma$}}(x)$ is a spin-1/2 vector operator. From this, the description in our work is easily established.

\renewcommand{\theequation}{B-\arabic{equation}}
\setcounter{equation}{0}  
\section*{APPENDIX B:  Resilience against phase-scrambling}  
\label{AppRes}

\begin{figure*}[t]
\psfig{figure=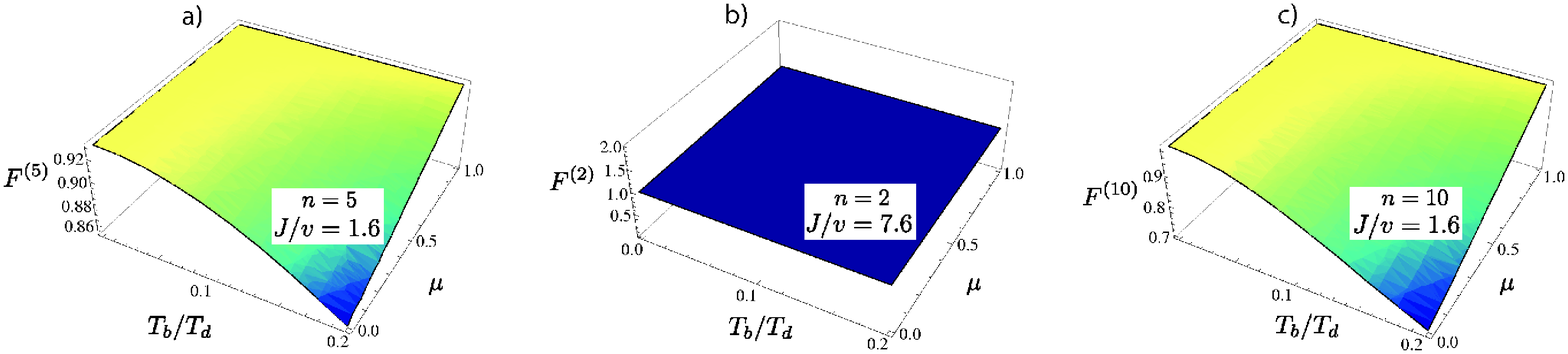,width=13.0cm,height=2.8cm}\\
\psfig{figure=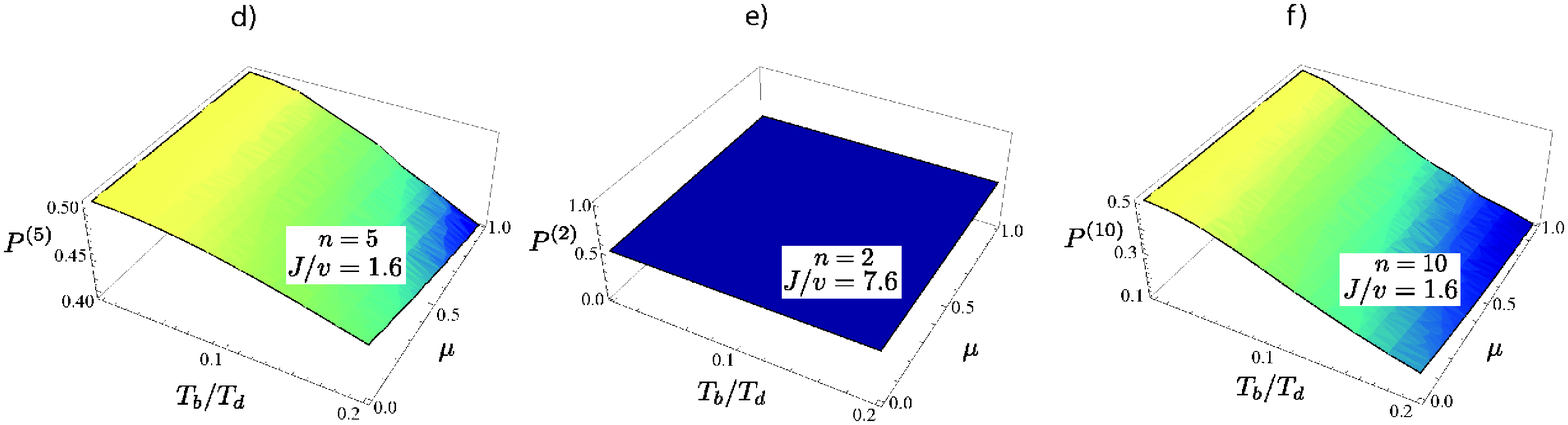,width=12.5cm,height=3.3cm}
\caption{Behavior against dephasing noise affecting the remote static spins $1$ and $2$ when model (3) is considered and for various settings of parameters $n$ and $J/v$. Panels (a), (b) and (c): state fidelity $F^{(n)}\!=\!_{12}\!\langle\Psi^{-}|\rho_{12}^{(n)}|\Psi^{-}\rangle_{12}$ for streams of unpolarized mediators (i.e. for $r=0$) against $T_b\!/\!T_d$ and $\mu$. Panels (d), (e) and (f): $P^{(n)}$ against $T_b\!/\!T_d$ and $\mu$.}  
\label{fig1SI}
\end{figure*}

In the case of the cavity-quantum electrodynamics implementation proposed in our paper, one should consider the effects of spurious energy-conserving interactions between the spin degree of freedom of each static particle and a background of phononic modes belonging to the substrate onto which each particle resides. The effects of such couplings can be effectively modelled as random phase-kicks over each spin's state occurring with a characteristic time $T_{d}$. Formally, they can be appropriately accounted for by standard quantum maps \cite{NC}. As stated in the body of the paper, a potential setup for testing our scheme is a GaN nanowire with embedded GaInN QDs. In this host, typical photonic wavelengths and group velocities  are $\lambda\!\sim\!$400 nm and $v_{ph}\!\sim\!c\!/$2, respectively. Under the conditions assumed in our study, each mediator takes a time $T_s\!\sim\!1/(v_{k_0}\Delta k)$ to be scattered off of spins 1 and 2 (see Ref.~\cite{rising-time} for details).  By taking $x_0\ug \pi/ k_0$ and $\Delta k/k_0\!\sim\! 10^{-2}$, so as to fulfill the RC, we obtain $T_s\sim10^{-14}$s, which is at least one order of magnitude smaller than the typical decoherence times in this setups. Therefore, in such regime phase kicks take place in the buffer time-window  $T_b$ between two successive scattering events. To test the resilience of  our protocol against $T_b\!/\!T_d$ we have performed a Monte Carlo numerical simulation of the proposed scheme. For a given number of steps $n$, the application of map $\mathcal{E}$ has been interspersed $n$ times with that describing phase kicks on 1 and 2, as coming from dephasing baths attached to each static spin. We explicitly allow for the possibility of correlations between the baths in the problem. The dephasing map depends on the dimensionless parameter \cite{NC} $T_b\!/\!T_d$ and the degree of noise correlation \cite{memoria} $\mu$. This quantity is the probability that 1 and 2 undergo correlated phase-kicks \cite{memoria} so that we have $\mu\ug0$ ($\mu\ug1$) for fully-uncorrelated (fully-correlated) baths. In Fig.~5 we study the resilience of the state fidelity against $T_b/T_{d}$ and $\mu$. Remarkably, for $\mu\ug1$ state fidelity is insensitive to dephasing noise. In fact, both the states entering the effective dynamical map responsible for the progressive projection onto the maximally entangled singlet state of the remote spins are known to be unaffected by perfectly correlated dephasing baths \cite{massimoProcRoy} (they are basically {\it decoherence-free} states for this class of noise). This is clearly reflected in Figs.~5(a), (b) and (c), where state fidelity is exactly $1$ for $\mu=1$, regardless of the ratio $T_b/T_d$. As the baths considered become only partially correlated, this striking and clear robustness is only partially spoiled: state fidelity remains very high even for a rather pessimistic $T_b/T_d$ and quite a large number $n$ of steps considered. We conclude by stressing that the spoiling effects of dephasing can be counterbalanced by the choice of a coupling strength large enough to achieve an effective projection onto the singlet state of the remote spins (with high fidelity) with only a single-mediator injection.

\end{document}